\documentclass[journal]{IEEEtran}
\usepackage{cite}
\usepackage{subfigure}
\usepackage{color}
\usepackage{graphicx}
\usepackage{dcolumn}
\usepackage{bm}
\usepackage{physics}
\usepackage{blindtext}

\usepackage{siunitx}

\usepackage{float}
\usepackage{graphicx,wrapfig}
\newcommand*\diff{\mathop{}\!\mathrm{d}}

\usepackage{color}
\usepackage[dvipsnames]{xcolor}

\definecolor{morange}{rgb}{0.8,0.2,0}
\definecolor{mblue}{rgb}{0,0.3,1.0}
\definecolor{mgreen}{rgb}{0.2,0.4,0}
\definecolor{mpurple}{rgb}{0.5 0.1 0.7}

\usepackage{stfloats}

\newcounter{storeeqcounter}
\newcounter{tempeqcounter}

\begin{document}

\title{ Analytical Investigation of Long-time Diffusion Dynamics in a Synaptic Channel with Glial Cells}

\author{Enes Oncu, Halil U. Ozdemir, Halil I. Orhan, Bayram Cevdet Akdeniz, Asaf Toprakci, Anil Aslihak,   \\H. Birkan Yilmaz, Ali E. Pusane, Tuna Tugcu, Fatih Dinc \vspace{-1em}
\thanks{© 2021 IEEE. Personal use of this material is permitted. Permission from IEEE must be obtained for all other uses, in any current or future media, including reprinting/republishing this material for advertising or promotional purposes, creating new collective works, for resale or redistribution to servers or lists, or reuse of any copyrighted component of this work in other works.}
}
\maketitle

\begin{abstract}
In this letter, we first derive the analytical channel impulse response for a cylindrical synaptic channel surrounded by glial cells and validate it with particle-based simulations. Afterwards, we provide an accurate analytical approximation for the long-time decay rate of the channel impulse response by employing Taylor expansion to the characteristic equations that determine the decay rates of the system. We validate our approximation by comparing it with the numerical decay rate obtained from the characteristic equation. Overall, we provide a fully analytical description for the long-time behavior of synaptic diffusion, e.g., the clean-up processes inside the channel after communication has long concluded. 
\end{abstract}

\begin{IEEEkeywords}
Diffusive synaptic communications, molecular communications, channel impulse response, synaptic channel \vspace{-1em}
\end{IEEEkeywords}

\IEEEpeerreviewmaketitle

\section{Introduction}
Molecular communications (MC) is a new communication paradigm that utilizes molecules as signal carriers for harsh environments or the living tissue~\cite{farsad2016ComprehensiveSO}. Molecular communications via diffusion (MCvD), an extensively studied topic in the literature, uses diffusion as the transport mechanism~\cite{
Jamali2018channelMF,wicke2018modelingDF}. To analyze the channel characteristics and the communication performance, channel impulse response (CIR) functions are derived for several simple topologies~\cite{wicke2018modelingDF,
yilmaz2014threeDC}, although not for all~\cite{
Jamali2018channelMF}. Deriving a CIR function is at the core of MC research since it enables the analytical evaluation of the performance of communications. Of particular interest for understanding noise-dominated long-time dynamics of CIR is the decay rate of the system, which provides a good approximation of the CIR for long times \cite{dincc2018impulse,lotter2020synaptic} and is useful for probing diffusive thermalization \cite{carslaw1992conduction}. Overall, current MCvD studies need to be enriched by means of deriving CIR or the decay rate of the CIR for various environments/channels. 

One of the inherent applications of MC is the synaptic communication that is the transmission of neurotransmitters inside the synaptic cleft \cite{khan2017diffusion}. As reviewed in \cite{gonzalez2019neuroplasticity}, several mental diseases can be related to deficiencies in synaptic processes. Therefore, understanding the dynamics of diffusion inside the synaptic channel can be beneficial for our grasp of mental disorders. While the synaptic process has already been extensively considered in the biology literature, for example see \cite{scimemi2009determining}, focusing on the molecular synaptic channel from a communication engineering perspective can pave the way for handling many open problems, as comprehensively explained in \cite{veletic2019synaptic}. One of these open problems is the analytical derivation of CIR of synaptic communication, which has been examined recently by omitting neurotransmitter degradation by enzymes and for a simplified geometry \cite{lotter2020synaptic,lotter2021saturating}.

In a recent work, \cite{lotter2020synaptic} provided a loose upper bound on the long-time decay rate of the CIR for a synaptic channel surrounded by glial cells and without neurotransmitter degradation inside the synaptic cleft. The channel considered in \cite{lotter2020synaptic} has the form of a square prism. While seemingly an arbitrary geometry choice, this is fully justified as the resulting upper bound is independent of channel geometry for all practical purposes. On the other hand, the assumption of zero degradation may be conflicting with the study of long-time dynamics\footnote{In most channels, decay rates are usually just shifted by the degradation rate. Thus, it is justified to initially ignore the degradation rate during analytical calculations. However, it has been shown in \cite{al2018modeling} that the degradation rate affects the decay rates in a complicated way, once the desorption mechanism is introduced to the system; thus, degradation cannot be considered separately.}. The main purpose of a degradation mechanism in communication channels is usually to \emph{clean-up the channel,} which can selectively reduce long-time noise drastically \cite{heren2015effect}. {Moreover, degradation is an integral part of the synaptic processes \cite{kandel2000principles}} and has to be considered in the derivation of the long-time decay rate. 

In this work, we consider the long-time dynamics of synaptic communication. While building on top of the previous work \cite{lotter2020synaptic}, we analytically investigate the long-time decay rate of the CIR for a cylindrical synaptic cleft {with an angular symmetry and homogeneous boundaries\footnote{While such assumptions might seem too simplifying, for example as to whether glial cells cover the whole boundary, these assumptions have been the driving force under the hood in both MC \cite{lotter2020synaptic} and neuroscience literature \cite{freche2011synapse} for modelling and gaining a good understanding for synaptic communication.}.} Specifically, our main contributions to the literature include:
\begin{itemize}
    \item We derive the analytical expression for the CIR of a cylindrical synaptic channel by including pre-synaptic and glial uptake, neurotransmitter-receptor binding, and unbinding, and neurotransmitter degradation when molecules are released from the center (later we discuss how to generalize for an arbitrary release location).
    \item We identify the characteristic equation for the decay rates of the CIR, which we then solve to obtain the decay rates numerically.
    \item We solve for the {long-time decay rate ($\lambda_{\rm decay}$)} analytically using a Taylor expansion, which is justified through an order of magnitude analysis of system parameters.
\end{itemize}
{ During a communication scenario, for the noise to die out quickly, one requires $\lambda_{\rm decay}^{-1}$ to be as small as possible. More concretely, calculating the decay rate provides us with a first order-of-magnitude check for the target channel communication time, as the sampling time should be picked larger than $ \lambda_{\rm decay}^{-1}$ to reduce inter-symbol interference (ISI). Thus, our main contribution, i.e., finding an accurate analytical approximation for the decay rate, can be immensely beneficial for the ISI mitigation studies in MC literature.}

We note that, while it is common practice to explore analytical asymptotic time dynamics of diffusion in the literature, \cite{carslaw1992conduction}, the molecule-receptor unbinding process we consider in this work leads to time-dependent boundary conditions and thus requires a more modern approach \cite{lotter2020synaptic,al2018modeling,deng2015modeling}. 

\vspace{-0.6em}

\section{System Model and Simulations}
In this letter, we consider the diffusion process inside a simplified cylindrical synaptic cleft, where the pre-synaptic cell is located at $z=0$, post-synapse is at $z=L$, and the whole system is surrounded by a glial cell located at $r=R_S$. There are several {components of the system model}, which we illustrate in Fig. \ref{fig:fig1} and discuss below:
\begin{figure}
    \centering
    \includegraphics[width=7cm]{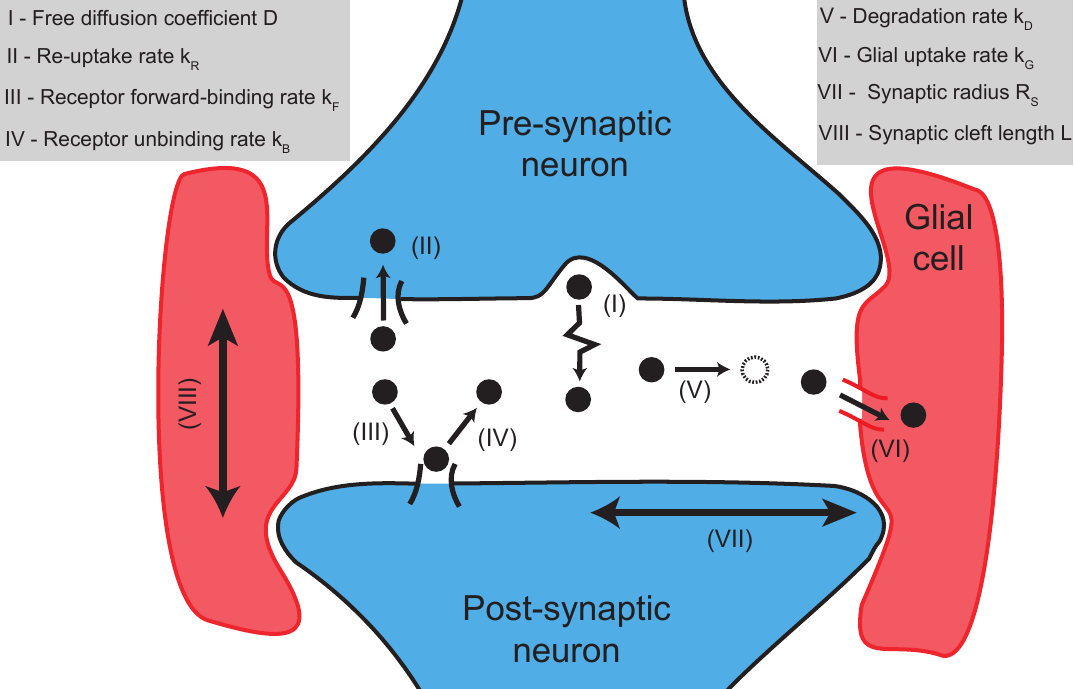}
    \caption{A simplified model of signal propagation inside the synaptic cleft.}
    \label{fig:fig1}
    \vspace{-1.5em}
\end{figure}

\emph{(I) Free diffusion:} Molecules inside the synaptic cleft diffuse with a diffusion coefficient of $D$. We simulate this process inside a single time step, $\Delta t$, by drawing samples from a normal distribution with mean $0$ and variance $2 D \Delta t $~\cite{yilmaz2014threeDC}. As an initial condition, we assume that $N$ molecules are released from the center of the pre-synapse.

\emph{(II) Re-uptake by pre-synapse:} Some molecules hitting the pre-synapse wall can be re-absorbed with some finite probability, given as $P_R = k_R \sqrt{\frac{\pi \Delta t}{D}}$, where $k_R$ is the pre-synaptic re-uptake rate. We simulate this process by drawing a random uniform number within $a \in [0,1]$ for each molecule of interest and absorb molecules when $a<P_R$, while we reflect the rest via a rollback mechanism~\cite{turan2018channel}.

\emph{(III) Forward-binding at post-synapse:} Diffusing neurotransmitters can bind with the receptors on the post-synapse, which we simulate similar to the re-uptake mechanism but with the probability $P_F = k_F \sqrt{\frac{\pi \Delta t}{D}}$, where $k_F$ is the post-synaptic forward binding rate.

\emph{(IV) Unbinding at post-synapse:} The receptor-neurotransmitter complex can unbind and release the transmitter back into the environment. We simulate this by first calculating the unbinding probability $P_B = 1-\exp(-k_B \Delta t)$, where $k_B$ is the post-synaptic unbinding (desorption) rate, and then drawing a random uniform number within $[0,1]$ for making the desorption decision. If the molecule is to be released back into the environment, we place the molecule back to its last position before being absorbed. 

\emph{(V) Degradation of neurotransmitters:} Diffusing neurotransmitters can degrade due to the enzymes present in the synaptic cleft with a probability $P_D \!=\! 1\!-\!\exp(-k_D \Delta t)$, where $k_D$ is the degradation rate. We probe this possibility at each step and remove the particles from the synaptic cleft if they degrade.

\emph{(VI) Glial uptake:} Diffusing neurotransmitters can be uptaken by the glial cells, which we simulate similar to the re-uptake mechanism, but with the probability $P_G = k_G \sqrt{\frac{\pi \Delta t}{D}}$, where $k_G$ is the glial uptake rate.

\emph{(VII-VIII) Spatial dimensions:} We assume that the synaptic {cleft is modeled by}
a cylindrical regime with synaptic radius $R_S$ and cleft length $L$. 

Within this work, the purpose of the particle-based simulations is to verify that our analytical derivations are accurate. In this letter, we simulate the boundary as a reactive surface, rather than a boundary with discrete receptors. Both are equivalent for our purposes and Lotter \emph{et al.} have already discussed the boundary homogenization assumption in \cite{lotter2020synaptic}.

\vspace{-0.6em}

\section{Analytical Derivation of CIR}

{In this section, we focus on finding the CIR, which is defined as the expected number of transmitter-receptor binding complex normalized by the number of released molecules $N$. As a start,} the diffusion of molecules is described by Fick's Law, given as
\begin{equation}
    \frac{\partial P_M(z,r,t)}{\partial t} = D \nabla^2 P_M(z,r,t) - k_D P_M(z,r,t). \label{eq:fick}
\end{equation}
Here, $P_M(z,r,t)$ is the probability density function for the messenger molecules and $\nabla^2$ is the Laplacian operator. The initial and boundary conditions can be written as
\begin{subequations} \label{eq:bound}
\begin{align}
    D \frac{\partial P_{M}(z,r,t)}{\partial z} \Big|_{z=0}&=k_R P_M(z\!=\!0,r,t), \label{eq:bound1} \\
    -D \frac{\partial P_{M}(z,r,t)}{\partial z} \Big|_{z=L}&= k_F P_M(z\!=\!L,r,t) \!+\! k_B h(t), \label{eq:bound2} \\
    -D\frac{\partial P_M(z,r,t)}{\partial r}\Big|_{r=R_S}&=k_G P_M(z,r\!=\!R_S,t), \label{eq:boundr} \\
    P_M(z,r,t=0) &= \frac{\delta(r)}{2\pi r} \delta(z), \label{eq:bound3}  
\end{align}
\end{subequations}
where $h(t)= D \int_{0}^t \frac{\partial P_{M}(z,r,\tau)}{\partial z}\Big|_{z=L} \diff \tau$. {Here, the first condition describes the re-uptake at the pre-synapse, the second one describes the forward reaction and the unbinding processes at the post-synapse, third one describes the uptake by glial cells, and the last one describes the initial release conditions of the molecules. For simplicity, we assume that the molecules are released from the center of the pre-synapse, as the long-time dynamics, e.g., the decay rate, are initial condition independent.} Considering these boundary conditions, we realize that the radial direction obeys the reactive boundary condition, whereas the axial direction has time-dependent components. Thus, we need to use a Laplace transform for the axial direction, following similar approaches before \cite{al2018modeling,lotter2020synaptic}. We start by assuming a separation of variables Ansatz of the form
\begin{equation}
    P_M(z,r,t) = \sum_n \kappa_n u_n(r) v_n(z,t),
\end{equation}
where $\kappa_n$ will later be chosen to satisfy the initial condition. Putting this Ansatz back into the equation and applying the radial boundary conditions, we find for the radial component~\cite{carslaw1992conduction}.
\begin{equation}
    u_n(r) = J_0\left( \alpha_n \frac{r}{R_S} \right), \; \text{with} \;\, J_1(\alpha_n) = \frac{R_S k_G}{D \alpha_n} J_0 (\alpha_n).
\end{equation}
Here, {$J_n(x)$ is the \emph{n}-th order Bessel function} of the first kind and $\{\alpha_n\}=\{\alpha_1,\alpha_2,\ldots\}$ is an ordered set of unitless eigenvalues {uniquely}  defined by the above characteristic equation such that $\alpha_1$ is the smallest eigenvalue. It is straightforward to show that $v_n(z,t)$ follows the 1D diffusion equation with
\begin{equation}
    \partial_t v_n(z,t) = D \partial_z^2 v_n(z,t)  - k_D^{(n)} v_n(z,t), 
\end{equation}
where the degradation rate of the 1D diffusing particles is shaped by a contribution coming from the glial cells $k_D^{(n)}=k_D+\frac{D}{R_S^2}\alpha_n^2$. Here, $\alpha_n$ quantitatively describes how strongly particles inside the mode $n$ interact with the glial cell. We take this analogy one step further and define $v_n(z,t=0)=\delta(z)$ such that $v_n(z,t)$ describes a particle diffusing in a 1D environment with degradation rate $k_D^{(n)}$. This leads to the interpretation that existence of glial cells shapes the signal in a way that is mathematically equivalent to releasing $n$ different particles with different degradation rates and shaping the final signal according to $\kappa_n$, which can be found using \eqref{eq:bound3} as
\begin{equation}
    \kappa_n = \frac{1}{\pi R_S^2 J_0 (\alpha_n)^2 \left(  1+ (\frac{R_S k_G}{D \alpha_n})^2 \right) }.
\end{equation}

Finally, we can find $v_n(z,t)$ by first performing a Laplace transform, then solving the problem, and finally transforming back via the inverse Laplace transform \cite{lotter2020synaptic,al2018modeling}, as given {at the bottom of this page} in (\ref{eq:vn}), with the definitions

\addtocounter{equation}{1}%
\setcounter{storeeqcounter}%
{\value{equation}}%

\begin{figure*}[!b]
\normalsize
\setcounter{tempeqcounter}{\value{equation}} 
\begin{IEEEeqnarray}{rCl}
\setcounter{equation}{\value{storeeqcounter}} 
\begin{aligned}
    v_n(z,t) = \frac{1}{L} \sum_{m=1}^\infty \frac{ \left(2 D \phi_{nm} \cos(\phi_{nm}) + 2 \gamma_{nm} L \sin(\phi_{nm}) \right) \cos(\phi_{nm} \frac{z}{L} + c_{nm}) \exp{- \lambda_{nm} t}}{\left( D\phi_{nm} \chi_{nm} \cos(\phi_{nm}+c_{nm}) + (D + \gamma_{nm} L \chi_{nm})\sin(\phi_{nm}+c_{nm}) + \psi_{nm} \cos(\phi_{nm}+c_{nm}) \right)}.
\end{aligned}
\label{eq:vn}
\end{IEEEeqnarray}
\setcounter{equation}{\value{tempeqcounter}} 
\hrulefill
\end{figure*}

\begin{subequations}
\begin{align}
   \gamma_{nm} &= \frac{\left(D \frac{\phi_{nm}^2}{L^2} +k_D^{(n)}\right) k_F}{D \frac{\phi_{nm}^2}{L^2} +k_D^{(n)} -k_B}, \\
   \chi_{nm} &= \left(1+\frac{D}{ k_R L} \sin^2c_{nm}\right), \\
   \psi_{nm} &= \frac{2 D k_B k_F \phi_{nm}}{ L \left(D \frac{\phi_{nm}^2}{L^2} +k_D^{(n)}-k_B \right)^2},\\
   c_{nm} &= \arctan(-\frac{k_R L}{D \phi_{nm}})
\end{align}
\end{subequations}
where $\phi_{nm}$ are the roots of the characteristic equation
\begin{align}
         \frac{\tan \phi_{nm} - \frac{k_R L}{D \phi_{nm}}}{\left( 1 +  \frac{k_R L}{D \phi_{nm}} \tan \phi_{nm}\right)} &= \gamma_{nm} \frac{L}{D \phi_{nm}} , \label{eq:rooteq}
\end{align}
and the decay rates are given as
\begin{equation} \label{eq:decay}
    \lambda_{nm} = D \frac{\phi_{nm}^2}{L^2} + k_D^{(n)} = D \frac{\phi_{nm}^2}{L^2} + k_D + D\frac{\alpha_n^2}{R_S^2}.
\end{equation}
{We note that $\lambda_{nm}$ can be recovered by applying the separation of variables Ansatz for $v_n(z,t)$ as well, without using the initial condition at all. Thus, the decay rate does not depend on the initial release location of molecules.}

Bringing all together, we find the probability density functions of diffusing molecules. Using this expression, we can find the channel impulse as
\begin{equation} \label{eq:cir}
    \text{CIR}(t)= -D \int_0^t \int_{0}^{R_S} 2 \pi r  \frac{\partial P_M(z,r,\tau)}{\partial z} \Big|_{z=L} \diff r  \diff \tau.
\end{equation}
The analytical expression for the CIR is used to validate the analytical results with the simulation results in Section~\ref{sec:results}. {Before wrapping up this section, we note that we have derived the CIR under the assumption that the molecules are released from the center. However, generalization to an arbitrary initial release location is straightforward by replacing $\kappa_n u_n(r)$ in the sum with the appropriate solution of the 2D closed disc diffusion problem corresponding to the initial release location, which is well-known in the diffusion literature \cite{carslaw1992conduction}. For example, for the initial condition where molecules are released uniformly from a circular shell with radius $r_0$, e.g. $P_M(z,r,t=0)=\delta(z)\frac{\delta(r-r_0)}{2\pi r}$, the full solution can be obtained by simply redefining $\kappa_n$ as $\kappa_n \times J_0(\alpha_n r_0/R_S)$.}

\vspace{-0.6em}

\section{Approximating Long-time Decay Rate of CIR}
For sufficiently large values of $t$, $\text{CIR}(t)$ decreases exponentially following the curve $ \exp(-\lambda_{\rm decay} t)$. We are interested in the smallest decay rate of the system ($\lambda_{\rm decay}$) that describes the dynamics in this region. This decay rate sets the effective time scale for how long the neurotransmitters will remain inside the synaptic cleft after the communication has been initiated. Biologically, {we expect this to be $\lambda_{\rm decay}^{-1} \sim O(\SI{}{\milli\second})$ \cite{scimemi2009determining}}. The long-time decay rate can be read-off from \eqref{eq:decay} as
\begin{equation} \label{eq:lambda}
    \lambda_{\rm decay} = D \frac{\phi_{11}^2}{L^2} + k_D + D\frac{\alpha_1^2}{R_S^2},
\end{equation}
where we pick $n=1$ and $m=1$ corresponding to the smallest decay rate, {as both $\phi_{nm}$ and $\alpha_n$ are by definition ordered sets}. At this point, we define the dimensionless system parameters that will provide further analytical insights, which is common practice in the literature \cite{carslaw1992conduction}:
\begin{align*}
    \tilde k_G = \frac{R_S}{D} k_G, \, \tilde k_{F/R} = \frac{L}{D} k_{F/R}, \, \tilde k_D^{(n)} = \frac{L^2}{D} k_D^{(n)}, \, \tilde k_B = \frac{L^2}{D} k_B.
\end{align*}
We have two key observations at this point: 
\begin{enumerate}
    \item All dimensionless parameters are much smaller than $1$ for our range of interest. 
    \item {For $\lambda_{\rm decay}^{-1} \sim O(\SI{}{\milli\second})$, which corresponds to the biological decay time of milliseconds \cite{scimemi2009determining}}, we require that $\phi_{11}\ll 1$ and $\alpha_1 < 1$. 
\end{enumerate}
The second one is a particularly robust observation, as $D$, $L$, and $R_S$ uniquely constrain $\phi_{11}$ and $\alpha_1$ to be small in (\ref{eq:lambda}) and their values are well-known within some intervals \cite{scimemi2009determining}. Thus, we can perform a Taylor approximation for the characteristic equations of $\alpha_n$ and $\phi_{nm}$ near zero to obtain
\begin{subequations}\label{eq:approx_alpha_phi}
\begin{align}
        \alpha_1 &\approx \sqrt{2 \tilde k_G},\\
        \phi_{11} & \approx \sqrt{\frac{ \tilde k_R \tilde k_B - (\tilde k_F+\tilde k_R)  \tilde k_D^{(1)} }{   \tilde k_B  +   \tilde k_R + \tilde k_F -\tilde k_D^{(1)}}},
\end{align}
\end{subequations}
where we ignore higher order multiplicative terms of $\tilde k_{X}$ for all possible $X$. At this point, we note that $\phi_{11}$ may not be necessarily real and could be imaginary for some cases. On the other hand, $\lambda_{\rm decay}$ is always real, following \eqref{eq:lambda}. Specifically, $\lambda_{\rm decay}$ is always positive by construction. Bringing \eqref{eq:lambda}, \eqref{eq:approx_alpha_phi}, and  $k_D^{(1)}=k_D+D\alpha_1^2/R_S^2\approx k_D + 2 k_G/R_S$ together, we approximate the long-time decay rate as
\begin{align} \label{eq:app}
\lambda_{\rm est} \!\approx\! \frac{k_B (k_R \!+\! k_D L \!+\! 2 k_G L/R_S) \!-\! L(k_D \!+\! 2k_G/R_S)^2 }{ k_R +  k_F + L (k_B-k_D-2k_G/R_S)}\!.
\end{align}
This Taylor approximation can lead to unreasonable results for some set of parameters, where one of the leading terms in the Taylor approximation vanishes. {This happens when $k_B$ is very low, where unbinding time becomes comparable to the other uptake processes.} Consequently, if $\lambda_{\rm est}$ is negative or unreasonably large, this means higher-order terms need to be taken into account. For the parameters we use in this work, this does not happen, so we are safe to use this approximation.

We note that \eqref{eq:app} is consistent with the upper bound in \cite[Eq. 38]{lotter2020synaptic}. By assumptions of the Taylor approximation $\phi_{11}\ll1$ so that $\frac{D}{L^2}\phi_{11}^2 \ll \frac{D}{L^2}$, which satisfies the diffusion  related part of the inequalities. For the reaction part, if we set $k_D \!=\! 0$ (as in \cite{lotter2020synaptic}) and $k_G=0$ (reflecting boundary), this reduces to $k_B\frac{k_R}{k_R+k_F + L k_B}< k_B\frac{k_R}{k_R+k_F}$, which satisfies the reaction related part of the inequality in \cite[Eq. 38]{lotter2020synaptic}. We will consider the more general case of the reactive-boundary in Section~\ref{sec:results}.

Before wrapping up this section, we note that \eqref{eq:rooteq}, which gives us the decay rates, can reproduce \cite[Eq. 22]{lotter2020synaptic} for the long-time behavior if we put the constraint on the degradation rate $k_D$ such that $k_D^{(1)}=C(\beta_1^2+\gamma_1^2)$, where $\beta_1$ and $\gamma_1$ are the smallest eigenvalues corresponding to the glial uptake in \cite[Eq. 22]{lotter2020synaptic} and $C$ is some constant that matches the units. Thus, findings of \cite{lotter2020synaptic} regarding long-time decay rate can be seen as a special case of our calculations despite geometric differences and the upper-bounds found in \cite[Eqs. 38 and 40]{lotter2020synaptic} are only relevant for a simplified version of the general problem.
\begin{table}
    \caption{Simulation parameters (consistent with \cite{lotter2020synaptic}).}
    \centering
    \begin{tabular}{|c|c|c|}
    \hline
        System parameter & Symbol & Default value \\
        \hline
        Diffusion coefficient & $D$ & $\SI{330}{\micro\meter^2 / \second}$\\
        \hline
        Synaptic cleft length & $L$ & $\SI{20}{\nano\meter}$ \\
        \hline
        Effective radius & $R_S$ & $\SI{150}{\nano\meter}$ \\
        \hline
        Post-synaptic forward binding rate & $k_F$ & $\SI{15}{\micro\meter / \second}$ \\
        \hline
        Pre-synaptic re-uptake rate & $k_R$ & $\SI{1.3}{\micro\meter / \second}$ \\
        \hline
        Glial uptake rate & $k_G$ & $\SI{26}{\micro\meter / \second}$ \\
        \hline
        Degradation rate & $k_D$ & $\SI{0.5}{\milli\second^{-1}}$ \\
        \hline
        Post-synaptic unbinding rate & $k_B$ & $\SI{8.5}{\milli\second^{-1}}$ \\
        \hline
        Simulation time step & $\Delta t$ & $\SI{10}{\nano\second}$ \\
        \hline
        Number of neurotransmitters & $N$ & $3000$ \\
        \hline
    \end{tabular}
    \vspace{-1em}
    \label{tab:1}
\end{table}

\begin{figure}[!tb]
    \centering
    \includegraphics[width=8.25cm]{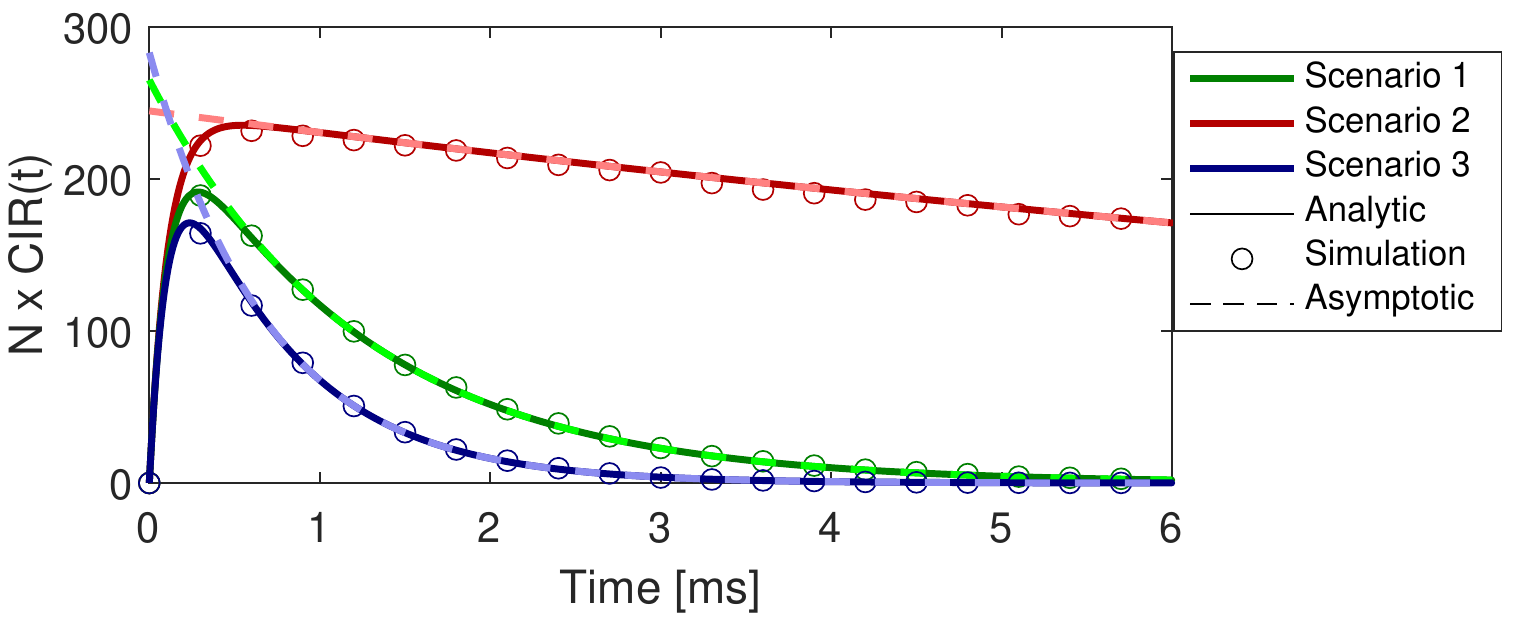}
    \vspace{-1em}
    \caption{Comparison of neurotransmitter-receptor complex at time $t$ for analytical and simulation results. For each sets of parameters, we run 100 simulations {with releasing} $N=3000$ molecules. For the analytical results, we provide both the numerical solution (solid) and the asymptotic solution (dashed) that has only one term corresponding to $\lambda_{\rm decay}$. Scenario 1 uses default parameters, {Scenario 2 sets $k_D\!=\!0$, and Scenario 3 sets $k_F\!=\!k_R=15 \mu m/s$ while keeping the rest as default value.}}
    \label{fig:fig2}
    \vspace{-1em}
\end{figure}

\vspace{-0.6em}
\section{Results}
\label{sec:results}

In this work, we consider the model parameters used in \cite{lotter2020synaptic}. Since our model also considers degradation and a cylindrical synaptic channel{ (with the aim of a more realistic scenario)}, we have two additional parameters compared to \cite{lotter2020synaptic}, $k_D$ and $R_S$. See Table \ref{tab:1} for the full set of model parameters. We pick $k_D$ and $R_S$ to be consistent with biological systems, although their exact numerical values are not of particular interest or relevance as our derivations are analytical.

We start by validating our analytical derivations with particle-based simulations. In Fig.~\ref{fig:fig2}, we plot the expected number of neurotransmitter-receptor binding complex, $N \times \text{CIR}(t)$, for three different scenarios with various system parameters. Moreover, we also plot an asymptotic analytic result, which takes only the leading term in the analytical expression for $\text{CIR}(t)$ in \eqref{eq:cir} and corresponds to the slowest long-time decay rate $\lambda_{\rm decay}$. Specifically, the asymptotic analytical result decays following $\exp(-\lambda_{\rm decay}t)$ and has been a particularly useful concept for calculating the approximate time duration of the communication \cite{dincc2018impulse}, which is $O\left(\lambda_{\rm decay}^{-1} \right)$ for a general system. As can be seen from the figure, analytical and numerical results agree well, whereas the asymptotic result perfectly describes the tail of the signal. We also note that Scenario 2 in Fig.~\ref{fig:fig2} reproduces one of the cases considered in \cite[Fig. 4]{lotter2020synaptic}.

\begin{figure}[!tb]
    \centering
    \includegraphics[width=9cm]{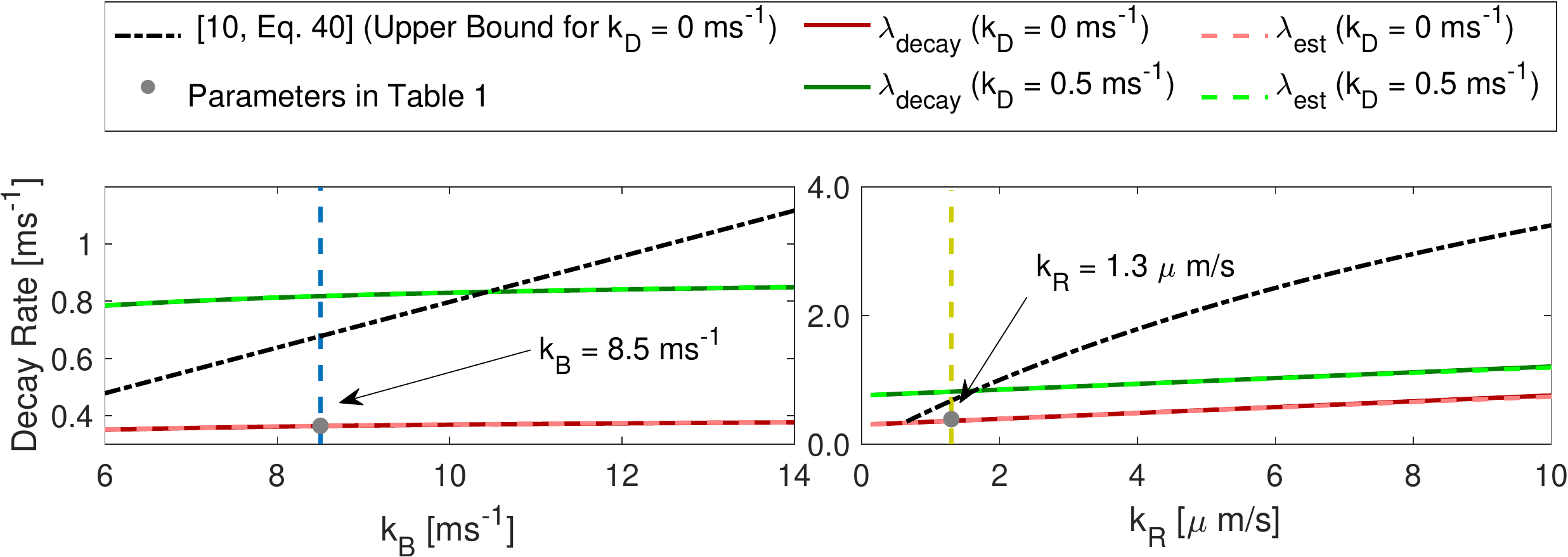}
    \caption{The comparison of the numerical $\lambda_{\rm decay}$ with {analytical estimation} $\lambda_{\rm est}$ and the upper bound found in \cite[Eq. 40]{lotter2020synaptic} for various system parameters. }
    \label{fig:fig3}
    \vspace{-1em}
\end{figure}

Having shown that the derived analytical result agrees with the particle-based simulations, we now turn our attention to the long-time behavior. In Fig. \ref{fig:fig3}, we illustrate how well $\lambda_{\rm decay}$ is approximated by {proposed} $\lambda_{\rm est}$ for various ranges of system parameters.  While we only illustrate variations in two system parameters due to the page limit, other parameters provide similar results. As can be seen from the figure, the approximation we derived in this letter works well within the parameters of interest. Moreover, the upper bound proposed by~\cite{lotter2020synaptic} is derived only for when $\phi_{11}$ is real, thus cannot be used for small re-uptake values (See Fig. \ref{fig:fig3}, right). However, small re-uptake values are of particular importance for re-uptake inhibition research~\cite{iversen2006neurotransmitter}.

{ Finally, we find that the Taylor approximation approach we used in this letter works well, without violating the condition $\phi_{11} \ll 1$, within at least an order of magnitude radius of the default parameters, picked in \cite{lotter2020synaptic} and in the literature cited therein. Only in cases, where one of the leading terms of the Taylor expansion vanishes, one is required to re-estimate (\ref{eq:app}) using higher-order terms. We find that this happens for unrealistically low $k_B$ values; otherwise the assumptions leading to (\ref{eq:app}) are quite robust due to biological constraints associated with observed spiking time dynamics and inferred diffusion parameters \cite{scimemi2009determining}.}

\vspace{-0.6em}

\section{Conclusion and Outlook}

In this work, we have considered the long-time dynamics of synaptic communication by finding an analytical approximation to the long-time decay rate of the system. We validated the approximation by comparing it with the numerical value calculated from the characteristic equation. 

While we have motivated the long-time decay rate through the channel impulse response, $\lambda_{\rm decay}$ provides a concrete description of the system's late-time dynamics. Noting that the probability density follows $P_M(z,r,t) \sim \exp(-\lambda_{\rm decay}t)$ for large $t$, all the observables, not only channel impulse response, have a long-time tail that is exponentially suppressed (at least) by $\lambda_{\rm decay}$. Thus, knowing $\lambda_{\rm decay}$ is equivalent to knowing when the system reaches equilibrium. 

Unlike many previous works, where the added value of analytical calculations on top of simulations is mostly for verification and/or finding bounds or semi-numerical tail approximations \cite{al2018modeling,lotter2020synaptic,lotter2021saturating,dincc2018impulse}, here a fully analytical description of the system provides us with a concrete symbolic understanding of the late-time diffusion dynamics. More concretely, previous work provided an upper bound that works for a certain range of parameters\cite{lotter2020synaptic}, here we have calculated an analytical approximation $\lambda_{\rm est}$ for a wide range of biologically plausible parameters. As future work, using the long-time decay rate, one can probe biology and/or design questions related to how each synaptic process contributes to the cleaning of the neurotransmitters from the synaptic cleft.

\vspace{-0.6em}



\begin{thebibliography}{10}
\providecommand{\url}[1]{#1}
\csname url@samestyle\endcsname
\providecommand{\newblock}{\relax}
\providecommand{\bibinfo}[2]{#2}
\providecommand{\BIBentrySTDinterwordspacing}{\spaceskip=0pt\relax}
\providecommand{\BIBentryALTinterwordstretchfactor}{4}
\providecommand{\BIBentryALTinterwordspacing}{\spaceskip=\fontdimen2\font plus
\BIBentryALTinterwordstretchfactor\fontdimen3\font minus
  \fontdimen4\font\relax}
\providecommand{\BIBforeignlanguage}[2]{{%
\expandafter\ifx\csname l@#1\endcsname\relax
\typeout{** WARNING: IEEEtran.bst: No hyphenation pattern has been}%
\typeout{** loaded for the language `#1'. Using the pattern for}%
\typeout{** the default language instead.}%
\else
\language=\csname l@#1\endcsname
\fi
#2}}
\providecommand{\BIBdecl}{\relax}
\BIBdecl

\bibitem{farsad2016ComprehensiveSO}
N.~{Farsad}, H.~B. {Yilmaz}, A.~{Eckford}, C.-B. {Chae}, and W.~{Guo}, ``A
  comprehensive survey of recent advancements in molecular communication,''
  \emph{IEEE Commun. Surveys Tuts.}, vol.~18, no.~3, pp. 1887--1919, 2016.

\bibitem{Jamali2018channelMF}
V.~{Jamali}, A.~{Ahmadzadeh}, W.~{Wicke}, A.~{Noel}, and R.~{Schober},
  ``Channel modeling for diffusive molecular communication—a tutorial
  review,'' \emph{Proc. of the IEEE}, vol. 107, no.~7, pp. 1256--1301, 2019.

\bibitem{wicke2018modelingDF}
W.~Wicke, T.~Schwering, A.~Ahmadzadeh, V.~Jamali, A.~Noel, and R.~Schober,
  ``Modeling duct flow for molecular communication,'' in \emph{Proc. of IEEE
  Global Commun. Conf. (GLOBECOM)}, 2018, pp. 206--212.

\bibitem{yilmaz2014threeDC}
H.~B. Yilmaz, A.~C. Heren, T.~Tugcu, and C.~Chae, ``Three-dimensional channel
  characteristics for molecular communications with an absorbing receiver,''
  \emph{IEEE Commun. Lett.}, vol.~18, no.~6, pp. 929--932, June 2014.

\bibitem{dincc2018impulse}
F.~Din{\c{c}}, B.~C. Akdeniz, A.~E. Pusane, and T.~Tugcu, ``Impulse response of
  the molecular diffusion channel with a spherical absorbing receiver and a
  spherical reflective boundary,'' \emph{IEEE Trans. Mol. Biol. Multi-Scale
  Commun.}, vol.~4, no.~2, pp. 118--122, 2018.

\bibitem{lotter2020synaptic}
S.~Lotter, A.~Ahmadzadeh, and R.~Schober, ``Synaptic channel modeling for dmc:
  Neurotransmitter uptake and spillover in the tripartite synapse,'' \emph{IEEE
  Trans. Commun.}, vol.~69, no.~3, pp. 1462--1479, 2021.

\bibitem{carslaw1992conduction}
H.~S. Carslaw and J.~C. Jaeger, \emph{Conduction of heat in solids}.\hskip 1em
  plus 0.5em minus 0.4em\relax Clarendon Press, 1959.

\bibitem{khan2017diffusion}
T.~Khan, B.~A. Bilgin, and O.~B. Akan, ``Diffusion-based model for synaptic
  molecular communication channel,'' \emph{IEEE Trans. NanoBiosci}, vol.~16,
  no.~4, pp. 299--308, 2017.

\bibitem{gonzalez2019neuroplasticity}
M.~P. Gonz{\'a}lez, A.~Macho-Gonz{\'a}lez, A.~Garcimartin, M.~E.
  L{\'o}pez-Oliva, J.~Benedi, and J.~J. Merino, ``Neuroplasticity and neuronal
  communications in the healthy and in the disease brain,'' \emph{J. Neurology,
  Neurological Sci. and Disorders}, vol.~5, no.~1, pp. 038--046, 2019.

\bibitem{scimemi2009determining}
A.~Scimemi and M.~Beato, ``Determining the neurotransmitter concentration
  profile at active synapses,'' \emph{Molecular Neurobiology}, vol.~40, no.~3,
  pp. 289--306, 2009.

\bibitem{veletic2019synaptic}
M.~Veleti{\'c} and I.~Balasingham, ``Synaptic communication engineering for
  future cognitive brain--machine interfaces,'' \emph{Proc. of the IEEE}, vol.
  107, no.~7, pp. 1425--1441, 2019.

\bibitem{lotter2021saturating}
S.~Lotter, M.~Sch{\"a}fer, J.~Zeitler, and R.~Schober, ``Saturating receiver
  and receptor competition in synaptic {DMC}: Deterministic and statistical
  signal models,'' \emph{arXiv preprint arXiv:2103.05341}, 2021.

\bibitem{al2018modeling}
M.~M. Al-Zu’bi and A.~S. Mohan, ``Modeling of ligand-receptor protein
  interaction in biodegradable spherical bounded biological
  micro-environments,'' \emph{IEEE Access}, vol.~6, pp. 25\,007--25\,018, 2018.

\bibitem{heren2015effect}
A.~C. Heren, H.~B. Yilmaz, C.-B. Chae, and T.~Tugcu, ``Effect of degradation in
  molecular communication: Impairment or enhancement?'' \emph{IEEE Trans. Mol.
  Biol. Multi-Scale Commun.}, vol.~1, no.~2, pp. 217--229, 2015.

\bibitem{kandel2000principles}
E.~R. Kandel, J.~H. Schwartz, T.~M. Jessell, S.~Siegelbaum, A.~J. Hudspeth, and
  S.~Mack, \emph{Principles of Neural Sci.}\hskip 1em plus 0.5em minus
  0.4em\relax McGraw-hill New York, 2000, vol.~4.

\bibitem{freche2011synapse}
D.~Freche, U.~Pannasch, N.~Rouach, and D.~Holcman, ``Synapse geometry and
  receptor dynamics modulate synaptic strength,'' \emph{PloS One}, vol.~6,
  no.~10, p. e25122, 2011.

\bibitem{deng2015modeling}
Y.~Deng, A.~Noel, M.~Elkashlan, A.~Nallanathan, and K.~C. Cheung, ``Modeling
  and simulation of molecular communication systems with a reversible
  adsorption receiver,'' \emph{IEEE Trans. Mol. Biol. Multi-Scale Commun.},
  vol.~1, no.~4, pp. 347--362, 2015.

\bibitem{turan2018channel}
M.~Turan, M.~S. Kuran, H.~B. Yilmaz, I.~Demirkol, and T.~Tugcu, ``Channel model
  of molecular communication via diffusion in a vessel-like environment
  considering a partially covering receiver,'' in \emph{IEEE Int. Black Sea
  Conf. on Commun. and Netw. (BlackSeaCom)}, 2018, pp. 1--5.

\bibitem{iversen2006neurotransmitter}
L.~Iversen, ``Neurotransmitter transporters and their impact on the development
  of psychopharmacology,'' \emph{British Journal of Pharmacology}, vol. 147,
  no.~S1, pp. S82--S88, 2006.

\end{thebibliography}
\end{document}